\documentclass[aps,pre,preprint,groupedaddress,epsfig,superscriptaddress,showpacs]{revtex4-1}
\usepackage{graphicx}
\usepackage{dcolumn}
\usepackage{bm}

\begin{document}


\title{Equilibrium sampling by re-weighting non-equilibrium simulation trajectories}


\author{Cheng Yang}
\affiliation{School of Physical Sciences, University of Chinese Academy of Sciences, Beijing 100049, China}

\author{Biao Wan}
\affiliation{School of Physical Sciences, University of Chinese Academy of Sciences, Beijing 100049, China}

\author{Shun Xu}
\affiliation{School of Physical Sciences, University of Chinese Academy of Sciences, Beijing 100049, China}

\author{Yanting Wang}
\affiliation{State Key Laboratory of Theoretical Physics, Institute of Theoretical Physics, Chinese Academy of Sciences, Beijing 100086, China}

\author{Xin Zhou\footnote{Author to whom correspondence should be addressed; Electronic Mail: xzhou@ucas.ac.cn  
}}
\affiliation{School of Physical Sciences, University of Chinese Academy of Sciences, Beijing 100049, China}


\date{\today}

\begin{abstract}
With the traditional equilibrium molecular simulations, it is usually difficult to efficiently visit the whole conformational space in complex systems, which are separated into some metastable conformational regions by high free energy barriers. The applied non-equilibrium process in simulations could enhance the transitions among these conformational regions, and the associated non-equilibrium effects can be removed by employing the Jarzynski equality (JE), then the global equilibrium distribution can be reproduced. However, the original JE requires the initial distribution of the non-equilibrium process is equilibrium, which largely limits the application of the non-equilibrium method in equilibrium sampling. By extending the previous method, the re-weighted ensemble dynamics (RED), which re-weights many equilibrium simulation trajectories from arbitrary initial distribution to reproduce the global equilibrium, to non-equilibrium simulations, we present a method, named as re-weighted non-equilibrium ensemble dynamics (RNED), to generalize the JE in the non-equilibrium trajectories started from an arbitrary initial distribution, thus provide an efficient method to reproduce the equilibrium distribution based on multiple independent (short) non-equilibrium trajectories. We have illustrated the validity of the RNED in a one-dimensional toy model and in a Lennard-Jones system to detect the liquid-solid phase coexistence.
\end{abstract}


\maketitle
\section{Introduction}
\label{I} 

Molecular dynamics (MD) simulation is an important tool for investigating macroscopic physical properties of systems by looking at their microscopic interactions. However, for many complex systems, such as biopolymers, multiple phase coexistence systems, etc., the energy surface is a very complex one in the very high dimensional conformational space. During a finite simulation time, the complicate conformational space could be reduced to several metastable states separated by high free energy barriers, thus a traditional MD simulation is locally trapped in one of the states for a long time, and the global equilibrium distribution is very difficult to reach. In the past decades, many enhanced sampling simulation techniques have been invented to circumvent that problem and great successes have been achieved in many cases  \cite{torrie1977nonphysical,Tsallis1988,swendsen1986replica,FrantzFD1990,LyubartsevMSV1992,MarinariP1992,Grubmueller1995,Voter1997b,WuW1999,YanP1999,MitsutakeSO2001,laio2002escaping,AndricioaeiDK2003,HamelbergSM2005b,ZhouJKZR2006,barducci2008well,wang2001efficient,gong2009structuring,ZhangM2009,XuZJ2015}. 
Nevertheless, more endeavours have been continuously put into improving the simulation efficiency for complex systems, such as proteins and DNAs, and for special thermodynamic situations, such as phase coexistences. 

Among several newly developed techniques fulfilling this purpose, one is the re-weighted ensemble dynamics (RED) method~\cite{gong2009structuring}. The RED generates many independent short simulation trajectories obeying the same dynamics but started from dispersed initial conformations. Due to the short simulation time, each of those trajectories can only visit a limited conformational region not far from its initial conformation. When lots of dispersed initial conformations are generated, the whole set of these trajectories could cover the whole important conformational region, but its conformational distribution is biased from the global equilibrium one, because these trajectories do not completely lose their memory of the initial conformations. The RED extracts the relations among these trajectories to establish a linear equation whose solution provides the weights of these trajectories for reproducing the global equilibrium. Practically, ones can use many independent computers to generate these trajectories simultaneously to shorten the waiting time for reaching the global equilibrium in comparison with traditional simulation techniques which usually generate a single long trajectory. 
In the RED, if each single simulation trajectory can visit a larger conformational region, less number of trajectories are required to cover the whole conformational space. Most of existing enhanced sampling techniques which bias the potential energy surface to improve the visiting efficiency of a single trajectory can be employed in the RED. 

In this paper, we show, non-equilibrium simulations under a special time-dependent Hamiltonian can be applied in the RED based on an extended microscopic form of Jarzynski equality (JE)~\cite{jarzynski1997nonequilibrium}, named as re-weighted non-equilibrium ensemble dynamics (RNED), then not only further improve sampling efficient of the RED, but greatly extend the application of the JE by removing its initial equilibrium requirement.

In the original JE, the initial conformational distribution of non-equilibrium trajectories is required to be equilibrium~\cite{PhysRevE.92.012131}, then the free energy difference between the initial and final system is related to the works of  these non-equilibrium trajectories. The JE is widely applied to estimate free energy in simulation and in single molecular experiments~\cite{park2003free,liphardt2002equilibrium}, but the requirement of initial equilibrium distribution usually limits its application in many complex cases where the initial equilibrium distribution itself is hard to get.

In principle, while the initial distribution is different from the equilibrium one, but the analytical form of deviation is known, we can apply the analytical deviation as re-weighting function to reproduce the initial equilibrium distribution, thus estimate the free energy difference, or reproduce the equilibrium distribution of the final system based on the microscopic form of JE presented by Hummer \emph{et al.}~\cite{hummer2001free}. However, the re-weighting technique is usually impractical due to too wide range of the re-weighting function, except in very low dimensional cases. 

In this paper, we show that the requirement of achieving a practical re-weighting to the initial distribution of non-equilibrium trajectories can be similarly reached in the RED frame, by constructing and solving the same linear equation in the RED from the simulation trajectories. 
 
Therefore, we present the RNED method by combining the RED and the JE to calculate the weights of non-equilibrium trajectories and to reproduce the global equilibrium distribution in both the initial and final systems. 

This paper is organized as follows. The basic theory of the RNED will be established in Sec.\ref{II}, the simulations and results will be introduced in Sec.\ref{III}, and a short conclusion will be given in Sec.\ref{IV}.

\section{Theory}
\label{II}
Let us consider an ensemble of simulation trajectories, $\{q_i(\tau)\},i=1,...,N$, started from different initial conformations, $\{q_i(0)\}$, under the same Hamiltonian $H(q;\lambda)$ with the time-dependent parameter $\lambda$. We assume $H(q,\lambda)$ equals to $H_0(q)$ while $0 \leq \tau \leq t_1$ and $t_2 \leq \tau \leq t$, \emph{i.e.}, two segments of equilibrium processes, but is time dependent between the two equilibrium segments, \emph{i.e.}, a non-equilibrium process in $(t_1, t_2)$. In the RED, we construct the equilibrium distribution by re-weighting the trajectories in a segment of equilibrium simulation, such as in the first equilibrium segment, $0 \leq \tau \leq t_1$, 
\begin{equation}
  \label{eq:2} 
  P_w = \frac{1}{\sum_k w_k} \sum_j w_jP_j(q)\rightarrow P_{eq}(q),
  \end{equation} 
with $N \rightarrow \infty.$ Here $w_k$ is the weight of the $k$th trajectory, and $P_k(q)$ is the distribution of its sample in conformational space $q$. The weight corresponds to the deviation of the initial distribution $P(q,0)=\frac{1}{N} \sum_j \delta(q-q_j(0))$ from the equilibrium distribution,   
\begin{equation}
  \label{eq:3}
  w_k = \frac{P_{eq}(q)}{P(q;0)} |_{q=q_k(0)}  \approx 
       \langle \frac{P_{eq}(q)}{P(q;0^+)} \rangle_{P_k(q,0^+)},
\end{equation}
where $\delta(\cdot)$ is the Dirac-$\delta$ function. $\langle \cdots \rangle_{P(q)}$ means the ensemble average under the distribution $P(q)$, which is estimated from the corresponding sample of the $P(q)$ in practice. 
Here we replace a single initial configuration at $\tau=0$ by a short initial segment of the trajectory $(\tau \in [0,0^+])$ to depress possible statistical errors~\cite{gong2009structuring}. It is easy to know $\sum_k w_k = N$ 
since $P(q;0^+)=\frac{1}{N}\sum_k P_k(q,0^+)$.

Substituting Eq.(\ref{eq:2}) into Eq.(\ref{eq:3}) leads to a linear equation of weights $\{w_j \}$, 
\begin{equation}
 \label{eq:linear-equation}
 \sum_j G_{ij} w_j = 0, 
\end{equation}
where $G_{ij} = \Lambda_{ij} - \delta_{ij}$ with $\Lambda_{ij} = \frac{1}{N} \langle \frac{P_j(q)}{P(q,0^+)} \rangle_{P_i(q,0^+)}$, and $\delta_{ij}$ is the kronecker $\delta$ symbol. By applying a complete set of orthonormalized basis functions $\{A^{\mu}(q)\}$~\cite{ChuanBiao}, we have~\cite{gong2009structuring,GongZO2015}
\begin{eqnarray}
\frac{P_j(q)}{P(q,0^+)} = \sum_{\mu}  A^{\mu}(q) \ a^{\mu}_j, 
\end{eqnarray}
with the expanded coefficient $a^{\mu}_j = \langle A^{\mu}(q) \rangle_{P_j(q)}$. 
Here the $\{A^{\mu}(q)\}$ is orthonormalized by a standard orthogon-normalization process from lots of (arbitrarily) chosen basis functions~\cite{ChuanBiao}, \emph{i.e.}, $\langle A^{\mu}(q) A^{\nu}(q) \rangle_{P(q,0^+)} = \delta_{\mu\nu}$. 
Consequently, 
\begin{eqnarray}
\Lambda_{ij}=\frac{1}{N}\sum_{\mu} \langle A^{\mu}(q) \rangle_{P_i(q,0^+)} \langle A^{\mu}(q) \rangle_{P_j(q)}.
\label{eq:expansion}
\end{eqnarray}
 
Eq.(\ref{eq:linear-equation}) is the key result of the RED~\cite{gong2009structuring}, whose matrix elements are estimated from simulation data, and whose solution provides $\{w_i\}$ for reproducing equilibrium properties. However, when the conformational space consists of multiple metastable regions separated by very high free energy barriers, simulation trajectories may be trapped in the local regions without crossing the barriers. In this case, Eq.(\ref{eq:linear-equation}) has multiple solutions, then the global equilibrium distribution can not be uniquely determined, which is the main limitation in application of the RED~\cite{gong2009structuring}.  
 
In the RNED, we design the non-equilibrium segment in $(t_1, t_2)$ to overcome the very high free energy barriers then to promote the transition events between metastable states, then build a connection between the equilibrium simulations before and after the non-equilibrium segment. The solution degeneration problem in Eq.(\ref{eq:linear-equation}) due to inadequate interstate transitions is then effectively overcome. We describe the remained details of the RNED below. 

Similar to Eq.(\ref{eq:2}), we can also construct the equilibrium distribution from the trajectories sampled in the second equilibrium segment, $t_2 \leq \tau \leq t$,
\begin{equation}
 \label{eq:4}
 P_w^{(2)} = \frac{1}{\sum_k w_k^{(2)}} \sum_j w_j^{(2)} P_j^{(2)}(q) \rightarrow P_{eq}(q).
\end{equation}
We can relate the weights $w_k^{(2)}$ to $w_k$ by the non-equilibrium work of the $k^{\mathrm{th}}$ trajectory in the non-equilibrium interval $(t_1,t_2)$~\cite{hummer2001free,hummer2005free}, 
 \begin{equation}
 \label{eq:5}
 w_k^{(2)} = w_k \Omega_k,
 \end{equation}
where $\Omega_k = \exp\{-W[q_k(\tau)]\}$, and the non-equilibrium work is 
\begin{equation}
 \label{eq:6}
 W[q_k(\tau)] = \int \limits_{\rm t_1}^{\rm t_2}\frac{\partial H(q_k(\tau);\lambda)}{\partial \lambda} \frac{d \lambda}{d \tau} d\tau.
 \end{equation} 
We can then combine the two weighted samples together to form the equilibrium distributions. For example, 
we estimate the equilibrium distribution by
 \begin{equation}
 \label{eq:7}
 P_{eq} \approx \frac{ P_w(q) + \gamma P_w^{(2)}(q)}{1 + \gamma}.
 \end{equation}
Here 
\begin{eqnarray}
\gamma=\frac{M_{eff}^{(2)}}{M_{eff}^{(1)}},
\end{eqnarray}
where $M_{eff}^{(2)}$ and $M_{eff}^{(1)}$ are the effective sizes of the weighted sample in the second equilibrium segment and the first equilibrium segment, respectively. 
The effective size of a weighted sample is usually smaller than its real size. The exact formula may be dependent on these weights besides the real size of sample, we might write~\cite{XuZJ2015}
\begin{eqnarray}
M_{eff} = \frac{(\sum w_i)^2} {\sum (w_i)^{2}} = M \frac{1}{1 + \sigma^{2}}. 
\end{eqnarray}
Here $M$ is the real size of sample, $\sigma$ is the fluctuation of the normalised weights, $\hat{w}_i = M \frac{w_{i}}{\sum w_i}$. 
Therefore, for unweighted sample, the effective size of sample is equal to its real size, but for weighted sample, the effective size is $1+\sigma^2$ times smaller than the real size $M$. 

While both the effective sizes of two samples are large sufficiently, any one of the two weighted samples itself gives a good estimate of the equilibrium distribution, thus any $\gamma$ could be applied in Eq.(\ref{eq:7}). In other word, the estimate is insensitive to the value of $\gamma$, although it exists a best $\gamma$ for the accuracy of estimate. We show that the selection of $\gamma$ is not sensitive to the reproduced equilibrium distribution in the RNED. Therefore, the exact formula of the effective size of sample and its weights is not important for our current purpose. 

We have
 \begin{equation}
 \label{eq:8}
 P_{eq} \approx \frac{1}{N (1+\gamma)}\sum_j w_j [P_j(q) + \frac{\gamma}{c} \Omega_j P_j^{(2)}(q)],
 \end{equation}
where $c \equiv \frac{\sum w_i \Omega_i}{\sum w_k}$ should ideally be unity according to the JE~\cite{jarzynski1997nonequilibrium}, but practically it might slightly differ from unity due to the statistical error caused by a finite value of $N$. We can then obtain the linear equation $\sum_j G_{ij} w_j =0$ with  
\begin{equation}
 \label{eq:9}
 G_{ij} = \frac{1}{1 + \gamma} [(\Lambda_{ij} - \delta_{ij}) + \gamma ( \frac{\Omega_j}{c} \Lambda_{ij}^{(2)} - \delta_{ij})],
\end{equation}
where $\Lambda_{ij}^{(2)} = \frac{1}{N} \langle \frac{P_j^{(2)}(q)}{P(q,0^+)} \rangle_{P_i(q,0^+)}$, same as the definition of $\Lambda_{ij}$ in Eq.~(\ref{eq:linear-equation}) except that $P_j(q)$ is replaced by $P_j^{(2)}(q)$. 
This equation can be rewritten as, 
\begin{eqnarray}
\sum_j \tilde{G}_{ij} w_j = 0, 
\label{eq:rned-equation}
\end{eqnarray}
where 
\begin{equation}
\label{eq:11}
\tilde{G}_{ij} = G_{ij} - \bar{G}_j = \frac{1}{1 + \gamma} ( {\tilde G}_{ij}^{(1)} + \gamma {\tilde G}_{ij}^{(2)} ),
\end{equation}
with ${\tilde G}_{ij}^{(1)} = \Lambda_{ij} - \delta_{ij}$ and ${\tilde G}_{ij}^{(2)} = \frac{\Omega_j}{c} (\Lambda_{ij}^{(2)} - \frac{1}{N})  - (\delta_{ij}-\frac{1}{N})$. 
Here $\bar{G_j} \equiv \frac{1}{N} \sum_i G_{ij}$. 
For convenience, we usually construct a symmetric matrix $\mathbf{H}=\tilde{G}^T \tilde{G}$  
(\emph{i.e.}, $H_{jk}=\sum_i \tilde{G}_{ij}\tilde{G}_{ik}$), and calculate the ground state of $\mathbf{H}$ with the weight $\mathbf{w} = (w_1,...,w_N)^T$, or equivalently, solve the equation
\begin{equation}
\label{eq:12}
\mathbf{H} \mathbf{w} = 0,
\end{equation}
to obtain the weight vector. In Eq.~(\ref{eq:12}), $\tilde{G}_{ij}$ depends on the parameters $\gamma$ and $c$. From the JE, $c = \frac{1}{N} \sum_i w_i \Omega_i$ is approximately equal to unity. 
  
\section{Results}
\label{III}
\subsection{One-dimensional potential}
We first employ a simple model with a single particle moving in a one-dimensional potential to illustrate the validity of the RNED method and what we should pay attention to when using this method. A particle moving in a one-dimensional potential $U(x)$ obeys the overdamped Langevin equation
    \begin{equation}
    \label{eq:13} 
    \frac{dx}{dt}=-\frac{dU}{dx}+\sqrt{2T}\xi(t),
    \end{equation}
where $T=0.2$ is the simulation temperature, $\xi(t)$ is a Gaussian white noise with zero mean and obeys correlation function $\langle \xi(t)\xi(t') \rangle=\delta(t-t')$. The integration time step is $0.001$, and
    \begin{equation}
    \label{eq:14}
    U(x)=x^4-kx^2,
    \end{equation}
where $k$ is a controllable variable which determines the energy surface. In our case we choose $k=3.2$ as the initial value and then change it over time to implement a non-equilibrium process. The two potential wells are located at $x=\pm \sqrt{\frac{k}{2}}$ and the height of the energy barrier is $\frac{k^2}{4}$. 
We deployed $1000$ trajectories with about $800$ started from the right well, and the others from the left. The set
    \begin{equation}
    \label{eq:15}    
    A^{\mu}(x)=\left\{
    \begin{array}{cc}
     1,\ & \ x_{\mu}<x<x_{\mu+1}, \\ 
     
     0,\ &\ {\rm others}.
    \end{array}
    \right.
    \\
    \end{equation}
was selected as the basis functions, where $x_{\mu}=-1.6 + 0.05 \mu$ with $\mu=1,...,64$ covering all the important regions of the conformational space. Each bin of $\{x_{\mu}\}$ can be combined with its neighbours if it contains too few samples.
       
\begin{figure}[H]
\centering
\includegraphics[width=0.6\textwidth, height=0.7\textwidth]{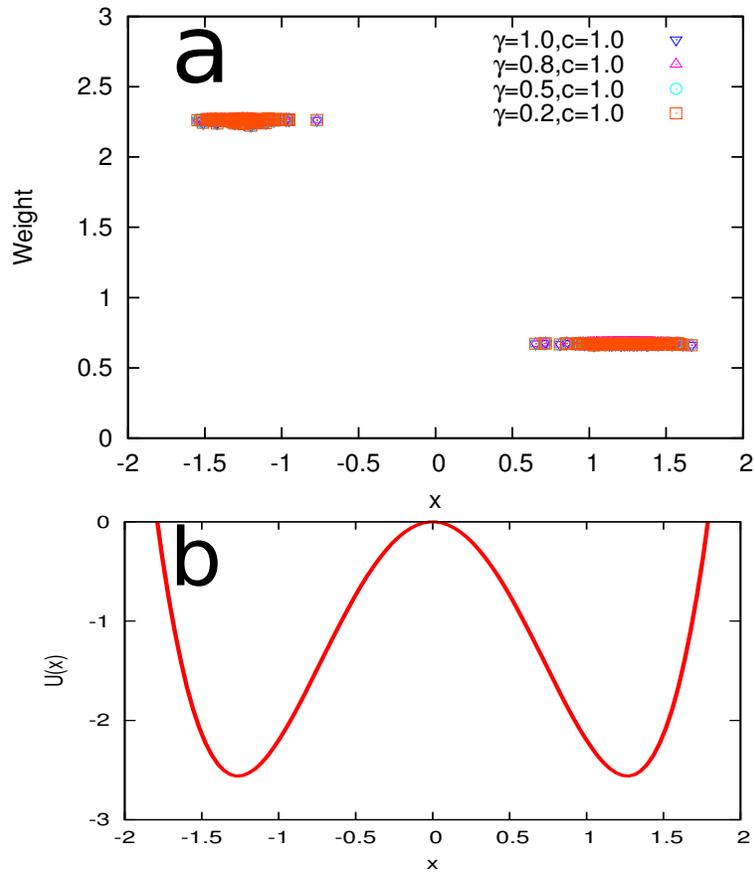}
\caption{\small (a) The relationship between the initial conformation of each trajectory and its weight. The difference between the weights with different $\gamma$ are very small. (b) The potential energy surface of $U(x)$.}
\label{Fig.1}
\end{figure}

During the simulation, the value of $k$ is changed with time as
\begin{equation}
    \label{eq:16}    
    k=\left\{
    \begin{array}{cc}
     3.2,\ & \ 0 \leq t<100,\\
     3.2-\Delta k\times int(\frac{t-100}{0.02}),\ & \ 100 \leq t<103,\\
     2.0,\ &\   \ 103 \leq t<153,\\
     2.0+\Delta k\times int(\frac{t-153}{0.02}),\ & \ 153 \leq t<156,\\
     3.2,\ & \ 156 \leq t \leq 256.\\
    \end{array}
    \right.
    \\
\end{equation}
where $\Delta k=0.008$ and the function $int(x)$ determines the largest integer smaller or equal to $x$. This non-equilibrium process is designed to first decrease the free energy barrier and then increase back to its original value. Samples are taken when $t \in [0,100]$ and $[156,256]$ with the interval of $\Delta t=0.1$. 
We choose $\gamma=1.0,0.8,0.5,0.2$ and set $c=1.0$ to calculate the weights of trajectories by the RNED, which are shown in Fig.\ref{Fig.1}a. We can see that the weights are independent of $\gamma$, so we fix $\gamma=1.0$ for simplicity in later calculations. Moreover, the trajectories in the same metastable state have similar weights, consistent with the RED method\cite{gong2009structuring}. The free energy surface is shown in Fig.\ref{Fig.1}b

\begin{figure}[H]
\centering
 \includegraphics[width=0.8\textwidth, height=0.6\textwidth]{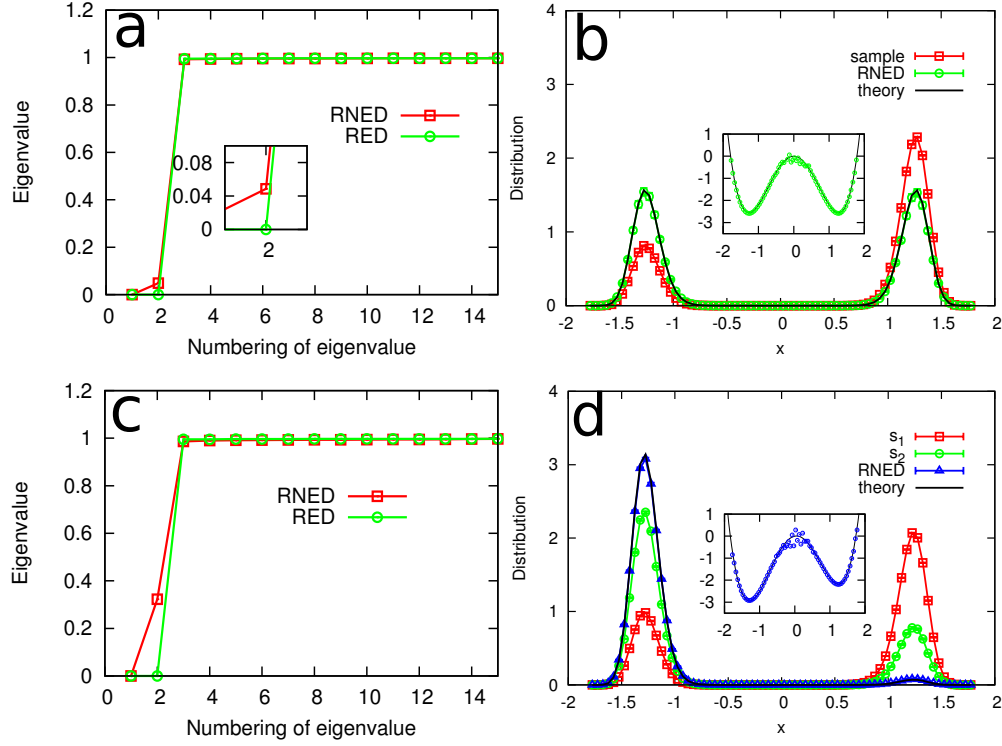}
\caption{\small (a) The first 15 smallest eigenvalues of RED and RNED when the potential energy is $U(x)$. The RED has two zero eigenvalues, which means there are few trajectories cross the free energy barrier, so we can't obtain the right weights via the RED in this case. (b) Compare the sampled distribution (red line) and weighted distribution (green line) of RNED with the theoretical distribution of $U(x)$ (black line). The weighted distribution is almost the same with theoretical distribution shows that the RNED method is effective. The inset shows the free energy surfaces of theory and obtained by the RNED. They are almost the same except on the free energy barrier for rare samples. (c) The first 15 smallest eigenvalues of RED and RNED when the potential energy is $U_b(x)$. RED method can't obtain weights of trajectories for there are two zero eigenvalues. (d) The black line is theoretical distribution of $U_b(x)$. The red line is the sampled distribution of first segment and the green line is the sampled distribution of second segment. Both sampled distributions are not equilibrium. The weighted distribution is consistent with theoretical distribution. The inset shows the energy surfaces of theory and obtained by the RNED. They are almost the same except on the free energy barrier for rare samples. }
\label{Fig.2}
\end{figure}

We have also performed a set of standard MD simulations with $k=3.2$. The initial conformations and total simulation time of each trajectory of this ensemble are the same as in the RNED method. The MD trajectories have been analysed with the RED method. The first $15$ smallest eigenvalues of the RED are shown in Fig.\ref{Fig.2}a with green line. Since the ground state is degenerate due to the fact that very few trajectories cross the free energy barrier, the weights cannot be determined uniquely. In the RNED, decreasing the free energy barrier helps the trajectories to transit between the two metastable states, so the ground state is non-degenerate (red line) and the elements of the corresponding eigenvector are the weights of the trajectories. Figure.\ref{Fig.2}b shows the distributions obtained by the RNED. The sampled distribution (red line) and the theoretical distribution (black line) are different, but the weighted distribution (green line) is almost the same with the latter, demonstrating the effectiveness of the RNED. The inset shows the free energy surface of theoretical (black line) and obtained by the RNED (green line). They are almost the same except on the free energy barrier for rare samples. In order to explain how RNED work further, we employ a asymmetric potential energy $U_b(x)=x^4-kx^2+0.3x$ as an additional example. The non-equilibrium process is the same as Eq.~(\ref{eq:16}) and standard MD simulations are also performed as a comparison. Fig.\ref{Fig.2}c shows the 15 smallest eigenvalues of RED (green) and RNED (red). RED method gets two zero eigenvalues while the RNED method is non-degenerate. The samples in RNED can be divided into two segments, the first segment is sampled before the non-equilibrium process and the second one is done after the non-equilibrium process. This two sampled distributions are shown in Fig.\ref{Fig.2}d (red and green) and both deviate the theoretical distribution of $U_b(x)$ (black). The weighted distribution is the blue line which is the same as theoretical distribution. The inset shows the free energy surface of theory (black) and obtained by RNED (blue).\\
 
\begin{figure}[H]
\centering
 \includegraphics[width=0.8\textwidth, height=0.6\textwidth]{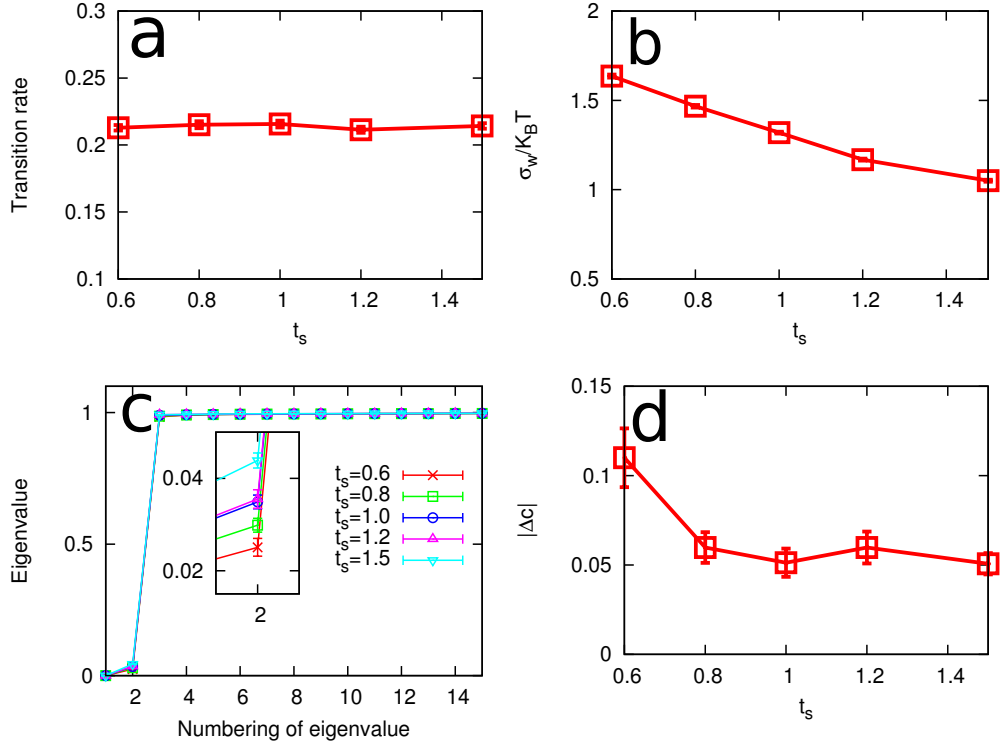}
\caption{\small Several parameters changed with $t_s$. (a) Transition rate between the two metastable states. (b) Standard deviation of the work. (c) The first $15$ smallest eigenvalues of the RNED. (d) The deviation of parameter $c$ from unity. In (a) (b) and (c) the statistical uncertainties are smaller than the symbols.}
\label{Fig.3}
\end{figure}

\begin{figure}[H]
\centering
 \includegraphics[width=0.8\textwidth, height=0.6\textwidth]{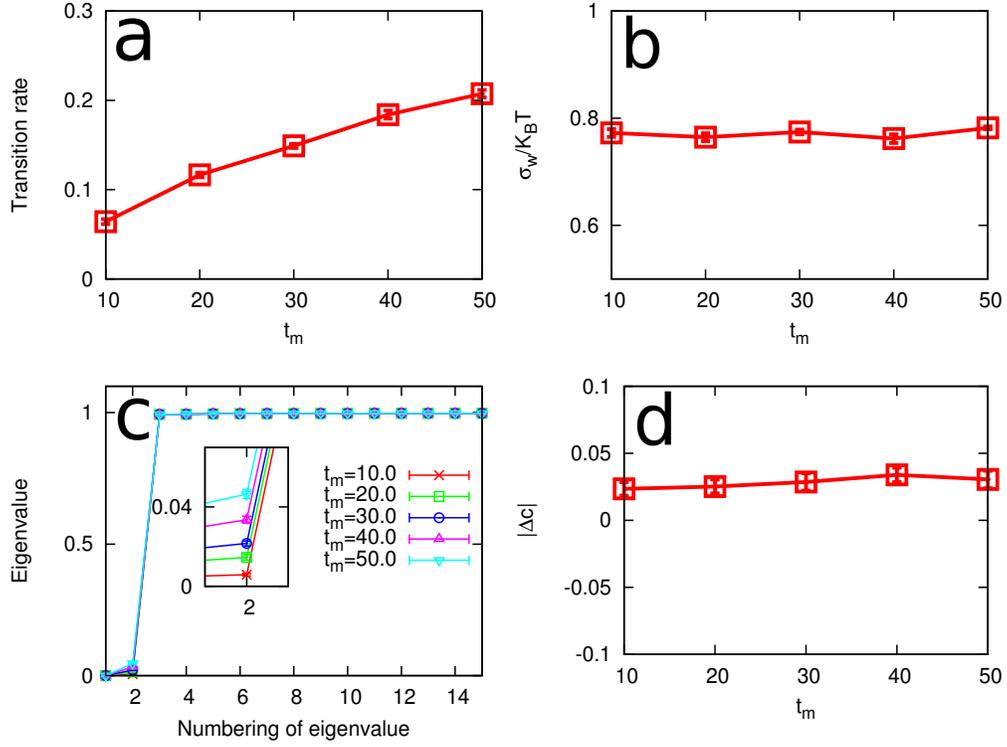}
\caption{\small Several parameters changed with $t_m$. (a) Transition rate between the two metastable states. (b) Standard deviation of the work. (c) The first $15$ smallest eigenvalues of the RNED. (d) The deviation of $c$ from unity. All statistical uncertainties are smaller than the symbols.}
\label{Fig.4}
\end{figure}

We then demonstrate what factors impact the results of the RNED. All the simulations are done under potential energy $U(x)$. Let $\Delta k=\frac{3.2-2.0}{m}$, where $m$ is a controllable parameter. $t_s=0.02m$ is the duration for $k$ to decrease from $3.2$ to $2.0$, and $t_m$ is the time for the system to stay at $k=2.0$. We first vary the switching time $t_s$, representing the speed of changing $k$, by choosing a different $m$ and keep $t_m=50$. The results are shown in Fig.\ref{Fig.3}. As $t_s$ increases, the transition rate (the ratio between the number of transition trajectories and total trajectories) does not change (see Fig.\ref{Fig.3}a). However, the standard deviation of the accumulated work decreases (see Fig.\ref{Fig.3}b). Correspondingly, the second smallest eigenvalue apparently rises when $t_s$ increases (see Fig.\ref{Fig.3}c). 
We employ $|\Delta c|=|c-1.0|$ to describe the deviation of $c$ from unity. It can be seen from Fig.\ref{Fig.3}d that the deviation decreases with $t_s$. Next, we vary $t_m$, the system evolving time with $k=2.0$, and keep $m=150$. The results are shown in Fig.\ref{Fig.4}. The transition rate increases along with $t_m$ (see Fig.\ref{Fig.4}a), but the standard deviation of the accumulated work does not change (see Fig.\ref{Fig.4}b). The second smallest eigenvalue becomes larger when $t_m$ increases (see Fig.\ref{Fig.4}c). In addition, the deviation of $c$ from unity is shown in Fig.\ref{Fig.4}d, which shows that $|\Delta c|$ dose not change with $t_m$.

The second smallest eigenvalue and $|\Delta c|$ are two indicators of the precision of the RNED. Weights are more precise when the second smallest eigenvalue deviates from zero more obviously and $|\Delta c|$ is closer to zero. The deviation between the second smallest eigenvalue and zero implies the ground state of the RNED is non-degenerate and $|\Delta c|$ is close to zero suggests JE is suitable for our example \cite{zuckerman2002theory,gore2003bias,park2004calculating}. Therefore, the transition rate and standard deviation of work are two main factors affecting the precision of the RNED. The RNED method can give a reasonable estimation of the equilibrium distribution only if both requirements are met.

\subsection{Lennard-Jones fluids}
\begin{figure}[H]
\centering
 \includegraphics[width=0.8\textwidth, height=0.6\textwidth]{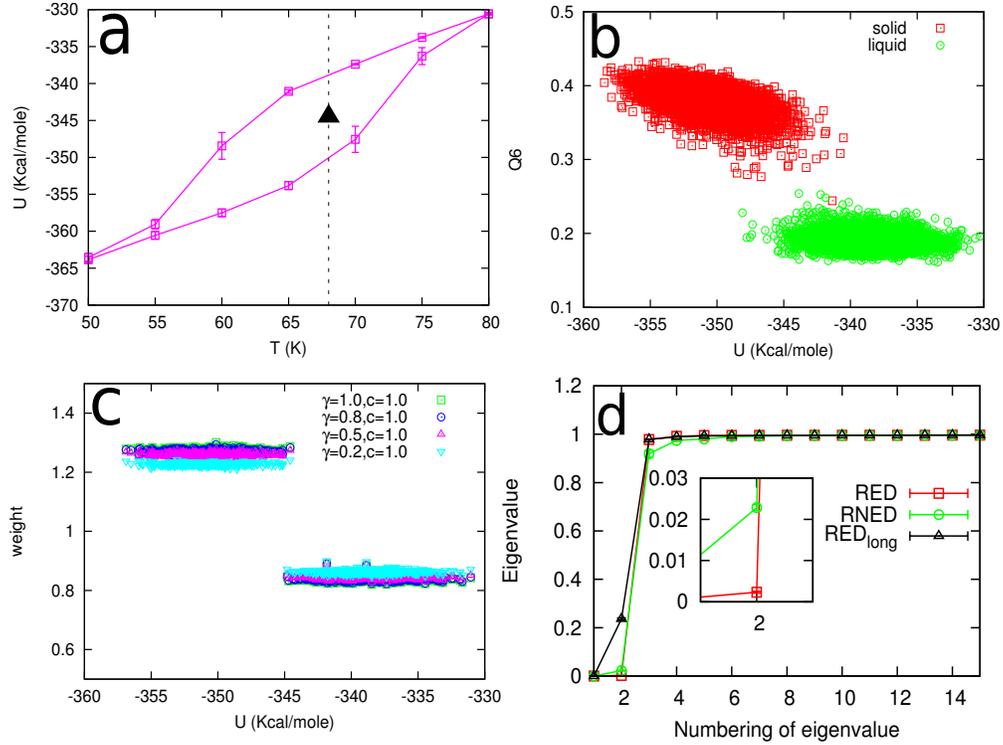}
\caption{\small (a) The hysteresis loop in the potential energy $U$ and temperature $T$ space. The dashed line is the temperature point at which we will reconstruct equilibrium distribution. The triangle is the equilibrium potential energy of L-J system we obtained by the RNED method. (b) ($Q6$, $P$) map for the metastable states of the L-J system, the red points are solid and the green points are liquid. (c) The relationship between the initial conformation of each trajectory and its weight. The difference between the weights with different $\gamma$ are very small. (d) The 15 smallest eigenvalues of RED and RNED. The black line is the result of RED for $200$ns-length ensemble. Red line is the result of RED for $8$ns-length ensemble. The second eigenvalue is very closed to zero implying there are few trajectories cross the free energy barrier. The green line is the result of RNED for non-equilibrium ensemble. The second eigenvalue deviates zero obviously when compare with red line.}
\label{Fig.5}
\end{figure}

Next we apply the RNED to a more complex system with the Lennard-Jones (L-J) potential. Our MD simulations of the system with the L-J potential are performed under the $NVT$ ensemble by using the LAMMPS simulation package \cite{lammps}. This system consists of $256$ particles and has a box size of $22.58{\textrm \AA} \times 22.58{\textrm \AA} \times 22.58{\textrm \AA}$ with the periodic boundary condition applied. The potential parameters of L-J are $\epsilon/k_{\rm B}=119.8$K and $\sigma=3.405{\textrm \AA}$ and the cut-off is $8.5{\textrm \AA}$. The properties of this model is similar to argon \cite{hansen1969phase}. As shown in Fig.\ref{Fig.5}a, there is a hysteresis loop in the potential energy $U$ and the temperature $T$ space while cooling and heating the system. Here we simulate $4$ ns at each temperature. 
The system locates in one of the two metastable states (liquid and solid) between $50$ K and $80$ K, depending on its history. The higher energy branch corresponds to the liquid state, while the lower energy branch corresponds to solid state. The dashed line represents $68$ K where we will reconstruct equilibrium distribution by the RNED and the triangle is the obtained equilibrium energy due to the RNED reconstruction. 
   
The L-J system is described by two order parameters. The first one is the potential energy $U$ and the second is the average local bond order parameter $Q_6=\langle Q_6(i) \rangle$ \cite{steinhardt1983bond,lechner2008accurate,russo2012microscopic}, where $\langle ..\rangle$ denotes averaging over all particles. $Q_6(i)$ is defined as 
\begin{eqnarray}
Q_6(i)=\sqrt{4\pi/13}|\hat{q}_6(i)|,
\end{eqnarray}
where 
\begin{eqnarray}
\hat{q}_{6m}(i)=\frac{1}{N_b(i)}\sum_{k=0}^{N_b(i)}q_{6m}(k), 
\end{eqnarray}
with $m = -6, -5 \cdots 5, 6$ and $N_b(i)$ the number of first neighbors around particle $i$, and 
\begin{eqnarray}
q_{6m}(k)=\frac{1}{N_b(k)}\sum_{j=1}^{N_b(k)}Y_{6m}(\hat{r}_{kj}). 
\end{eqnarray}
Here $Y_{6m}$ is the spherical harmonic function, $\hat{r}_{kj}$ is the normalized vector from particle $k$ to particle $j$. The two metastable states of L-J system in ($U$,$Q_6$) map are shown in Fig.\ref{Fig.5}b, the red points are solid conformations and green points are liquid conformations \cite{lechner2008accurate,russo2012microscopic}.

The functions of $U$ and $Q_6$ 
   \begin{equation} \small
   \label{eq:17}
    A^{\mu}(U,Q_6)=\left\{
    \begin{array}{cc}
     1,\ & \ U^l<U<U^{l+1}, Q_6^k<Q_6<Q_6^{k+1},\\   
     0,\ &\ {\rm others}.
    \end{array}
    \right.
    \end{equation}
are selected as the basis functions, where $U^l=-360+l,l=1,..,30$ and $Q_6^k=0.1+0.02k,k=1,..,20,$ which cover the most important part of the conformational space.

We have simulated $1000$ non-equilibrium trajectories and each trajectory lasts $8.0$ns. There are $500$ trajectories starting from solid conformations and others starting from liquid conformations. The non-equilibrium process was implemented 
by changing the potential energy to $U_{eff}=U+\frac{\alpha(t)}{2\beta}(U-U_0)^2$, where $U$ is the physical potential energy of the L-J system,
$\beta=1/k_{\rm B}T$ with $k_{\rm B}$ the Boltzmann factor, and $\alpha$ is a controllable parameter. When $\alpha=0.0$, it degenerates into a standard MD simulation. $U_0$ is chosen to approach to the position of the free energy barrier between liquid and solid, about $U_0 = - 345.0$ in our case. The system evolves under the new potential energy $U_{eff}$, thus the free energy surface is changed, leads a higher transition rate between the liquid and solid states. $\alpha$ changes with time as
\begin{equation}
    \label{eq:18}    
    \alpha=\left\{
    \begin{array}{cc}
     0.0,\ & \ 0.0 \leq t<0.5,\\
     0.0+0.0025\times int(\frac{t-0.5}{0.005}),\ & \ 0.5 \leq t<1.5,\\
     0.5,\ &\   \ 1.5 \leq t<6.5,\\
     0.5-0.0025\times int(\frac{t-6.5}{0.005}),\ & \ 6.5 \leq t<7.5,\\
     0.0,\ & \ 7.5 \leq t \leq 8.0.\\
    \end{array}
    \right.
    \\
    \end{equation}
Samples are taken when $t\in [0,0.5]$ and $[7.5,8.0]$. We choose $\gamma=1.0,0.8,0.5,0.2$ and set $c=1.0$ to calculate the weights of non-equilibrium ensemble with the RNED method, as shown in Fig.\ref{Fig.5}c. Since the weights are independent of the specific choice of $\gamma$, we set $\gamma=1.0$ for simplicity. The 15 smallest eigenvalues of RNED is shown in Fig.\ref{Fig.5}d (green line). We also simulated $1000$ equilibrium trajectories as comparison. Each trajectory of comparison ensemble lasted $8.0$ ns,too. Then we used the RED method to analyse this ensemble. The first $15$ smallest eigenvalues are show in Fig.\ref{Fig.5}d (red line). The second eigenvalue is very closed to zero when compare with the RNED's for there are few trajectories cross the free energy barrier. So the precise weights of $8.0$ns-length ensemble can't be obtained by the RED method. As addition, we simulated $200$ equilibrium trajectories with $100$ trajectories starting from solid state and the others starting from liquid state. Each trajectories lasted $200$ ns. We analyse this $200$ns-length ensemble with the RED method. The first $15$ smallest eigenvalues are show in Fig.\ref{Fig.5}d (black line). The second eigenvalue deviates zero obviously.\\

\begin{figure}[H]
\centering
 \includegraphics[width=0.8\textwidth, height=0.6\textwidth]{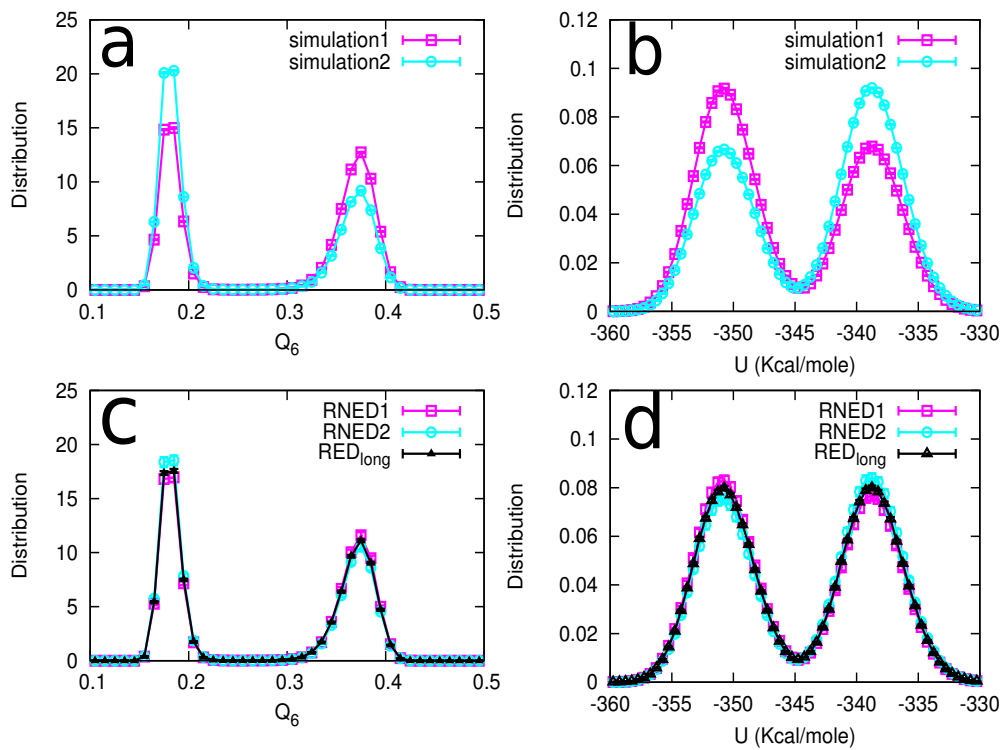}
\caption{\small (a) The sampled distributions of simulation1 and simulation2 in parameter $U$ space. (b) The sampled distributions of simulation1 and simulation2 in $Q_6$ space. (c) (d) The weighted distributions of these tow non-equilibrium ensembles. Black lines are the weighted distributions of $200$ns-length ensemble obtained by the RED method.}
\label{Fig.6}
\end{figure}

We discarded $100$ trajectories starting from liquid state of non-equilibrium ensemble randomly and the others constitute a new ensemble which we call $simulation1$. In the same way, we discarded $200$ trajectories starting from solid state of non-equilibrium ensemble and we call the rest of non-equilibrium ensemble $simulation2$. The sampled distributions of these two ensemble in different order parameter space are shown in Fig.\ref{Fig.6}a and Fig.\ref{Fig.6}b. Then we use RNED method to analyse these ensembles. The weighted distributions of two ensembles in parameter $Q_6$ space are shown in Fig.\ref{Fig.6}c and in parameter $U$ space are shown in Fig.\ref{Fig.6}d. The weighted distributions of two ensembles are very similar in both order parameter space. The black lines in Fig.\ref{Fig.6}c and Fig.\ref{Fig.6}d are weighted distributions of $200$ns-length ensemble obtained by RED method. The RNED method can give the same weighted distribution from different initial conformations and the weighted distribution is also consistent with the weighted distribution of longer MD ensemble analysed by the RED method. 

\section{Conclusions}
\label{IV}
In this work, we have generalized the RED scheme to be the RNED method with the help of the Jarzynski Equality. The RNED method is especially useful when the 
free energy barrier is so high that transition events are not adequate. 
The designed non-equilibrium process greatly enhances the transition rate between different free energy basins and the RNED method provides the weights for such trajectories systematically to reconstruct global equilibrium properties. 
This novel method has been successfully applied to two systems. For the one-dimensional system, we have compared RED and RNED 
with exactly the same simulation time and 
initial conformations. The results show that the RNED method is more efficient and the non-equilibrium work and the number of transition events influence the accuracy of the RNED. For the L-J system, we have calculated the equilibrium distribution by  RNED method started from different initial distributions and the equilibrium distributions are consistent with the weighted distribution of longer MD simulation obtained by the RED method. 

The RNED method is advantageous in the sense that it does not require much \emph{a priori} knowledge about the simulated system.  
We can do some short simulations at different conditions, such as temperatures or pressures, to obtain the initial states of non-equilibrium trajectories. The non-equilibrium process can be designed in many different ways depending on the studied system and problem. For instance, we can scale the Hamiltonian in the order parameter space or pull part of the system with an external force. The RNED method works as long as the non-equilibrium trajectories can satisfy our criteria discussed in Sec.\ref{III}.

\acknowledgements{ 
This work was supported by NSFC under Grant No. 11175250 and the Open Project from State Key Laboratory of Theoretical Physics. The authors thank  fruitful discussions with D. P. Landau. X.Z. thanks the financial support of the Hundred Talent Program of the Chinese Academy of Sciences. }

 
 \bibliographystyle{apsrev4-1}
 \bibliography{ref.bib} 

\begin{thebibliography}{10}%
\makeatletter
\providecommand \@ifxundefined [1]{%
 \ifx #1\undefined \expandafter \@firstoftwo
 \else \expandafter \@secondoftwo
\fi
}%
\providecommand \@ifnum [1]{%
 \ifnum #1\expandafter \@firstoftwo
 \else \expandafter \@secondoftwo
\fi
}%
\providecommand \enquote [1]{``#1''}%
\providecommand \bibnamefont  [1]{#1}%
\providecommand \bibfnamefont [1]{#1}%
\providecommand \citenamefont [1]{#1}%
\providecommand\href[0]{\@sanitize\@href}%
\providecommand\@href[1]{\endgroup\@@startlink{#1}\endgroup\@@href}%
\providecommand\@@href[1]{#1\@@endlink}%
\providecommand \@sanitize [0]{\begingroup\catcode`\&12\catcode`\#12\relax}%
\@ifxundefined \pdfoutput {\@firstoftwo}{%
 \@ifnum{\z@=\pdfoutput}{\@firstoftwo}{\@secondoftwo}%
}{%
 \providecommand\@@startlink[1]{\leavevmode\special{html:<a href="#1">}}%
 \providecommand\@@endlink[0]{\special{html:</a>}}%
}{%
 \providecommand\@@startlink[1]{%
  \leavevmode
  \pdfstartlink
   attr{/Border[0 0 1 ]/H/I/C[0 1 1]}%
   user{/Subtype/Link/A<</Type/Action/S/URI/URI(#1)>>}%
  \relax
 }%
 \providecommand\@@endlink[0]{\pdfendlink}%
}%
\providecommand \url  [0]{\begingroup\@sanitize \@url }%
\providecommand \@url [1]{\endgroup\@href {#1}{\urlprefix}}%
\providecommand \urlprefix [0]{URL }%
\providecommand \Eprint[0]{\href }%
\@ifxundefined \urlstyle {%
  \providecommand \doi [1]{doi:\discretionary{}{}{}#1}%
}{%
  \providecommand \doi [0]{doi:\discretionary{}{}{}\begingroup
  \urlstyle{rm}\Url }%
}%
\providecommand \doibase [0]{http://dx.doi.org/}%
\providecommand \Doi[1]{\href{\doibase#1}}%
\providecommand \bibAnnote [3]{%
  \BibitemShut{#1}%
  \begin{quotation}\noindent
    \textsc{Key:}\ #2\\\textsc{Annotation:}\ #3%
  \end{quotation}%
}%
\providecommand \bibAnnoteFile [2]{%
  \IfFileExists{#2}{\bibAnnote {#1} {#2} {\input{#2}}}{}%
}%
\providecommand \typeout [0]{\immediate \write \m@ne }%
\providecommand \selectlanguage [0]{\@gobble}%
\providecommand \bibinfo [0]{\@secondoftwo}%
\providecommand \bibfield [0]{\@secondoftwo}%
\providecommand \translation [1]{[#1]}%
\providecommand \BibitemOpen[0]{}%
\providecommand \bibitemStop [0]{}%
\providecommand \bibitemNoStop [0]{.\EOS\space}%
\providecommand \EOS [0]{\spacefactor3000\relax}%
\providecommand \BibitemShut [1]{\csname bibitem#1\endcsname}%
\bibitem{torrie1977nonphysical}%
  \BibitemOpen
  \bibfield{author}{%
  \bibinfo {author} {\bibfnamefont{G.~M.}\ \bibnamefont{Torrie}}\ and\ \bibinfo
  {author} {\bibfnamefont{J.~P.}\ \bibnamefont{Valleau}},\ }%
  \bibfield{journal}{%
  \bibinfo {journal} {J. Comput. Phys.}\ }%
  \textbf{\bibinfo {volume} {23}},\ \bibinfo {pages} {187} (\bibinfo {year}
  {1977})%
  \bibAnnoteFile{NoStop}{torrie1977nonphysical}%
\bibitem{Tsallis1988}%
  \BibitemOpen
  \bibfield{author}{%
  \bibinfo {author} {\bibfnamefont{C.}~\bibnamefont{Tsallis}},\ }%
  \bibfield{journal}{%
  \bibinfo {journal} {J. Stat. Phys.}\ }%
  \textbf{\bibinfo {volume} {52}},\ \bibinfo {pages} {479} (\bibinfo {year}
  {1988})%
  \bibAnnoteFile{NoStop}{Tsallis1988}%
\bibitem{swendsen1986replica}%
  \BibitemOpen
  \bibfield{author}{%
  \bibinfo {author} {\bibfnamefont{R.~H.}\ \bibnamefont{Swendsen}}\ and\
  \bibinfo {author} {\bibfnamefont{J.-S.}\ \bibnamefont{Wang}},\ }%
  \bibfield{journal}{%
  \bibinfo {journal} {Phys. Rev. Lett.}\ }%
  \textbf{\bibinfo {volume} {57}},\ \bibinfo {pages} {2607} (\bibinfo {year}
  {1986})%
  \bibAnnoteFile{NoStop}{swendsen1986replica}%
\bibitem{FrantzFD1990}%
  \BibitemOpen
  \bibfield{author}{%
  \bibinfo {author} {\bibfnamefont{D.~D.}\ \bibnamefont{Frantz}}, \bibinfo
  {author} {\bibfnamefont{D.~L.}\ \bibnamefont{Freeman}},\ and\ \bibinfo
  {author} {\bibfnamefont{J.~D.}\ \bibnamefont{Doll}},\ }%
  \bibfield{journal}{%
  \bibinfo {journal} {J. Chem. Phys.}\ }%
  \textbf{\bibinfo {volume} {93}},\ \bibinfo {pages} {2769} (\bibinfo {year}
  {1990})%
  \bibAnnoteFile{NoStop}{FrantzFD1990}%
\bibitem{LyubartsevMSV1992}%
  \BibitemOpen
  \bibfield{author}{%
  \bibinfo {author} {\bibfnamefont{A.~P.}\ \bibnamefont{Lyubartsev}}, \bibinfo
  {author} {\bibfnamefont{A.~A.}\ \bibnamefont{Martsinovski}}, \bibinfo
  {author} {\bibfnamefont{S.~V.}\ \bibnamefont{Shevkunov}},\ and\ \bibinfo
  {author} {\bibfnamefont{P.~N.}\ \bibnamefont{Vorontsov-Velyaminov}},\ }%
  \bibfield{journal}{%
  \bibinfo {journal} {J. Chem. Phys.}\ }%
  \textbf{\bibinfo {volume} {96}},\ \bibinfo {pages} {1776} (\bibinfo {year}
  {1992})%
  \bibAnnoteFile{NoStop}{LyubartsevMSV1992}%
\bibitem{MarinariP1992}%
  \BibitemOpen
  \bibfield{author}{%
  \bibinfo {author} {\bibfnamefont{E.}~\bibnamefont{Marinari}}\ and\ \bibinfo
  {author} {\bibfnamefont{G.}~\bibnamefont{Parisi}},\ }%
  \bibfield{journal}{%
  \bibinfo {journal} {Europhys. Lett.}\ }%
  \textbf{\bibinfo {volume} {19}},\ \bibinfo {pages} {451} (\bibinfo {year}
  {1992})%
  \bibAnnoteFile{NoStop}{MarinariP1992}%
\bibitem{Grubmueller1995}%
  \BibitemOpen
  \bibfield{author}{%
  \bibinfo {author} {\bibfnamefont{H.}~\bibnamefont{Grubmueller}},\ }%
  \bibfield{journal}{%
  \bibinfo {journal} {Phys. Rev. E}\ }%
  \textbf{\bibinfo {volume} {52}},\ \bibinfo {pages} {2893} (\bibinfo {year}
  {1995})%
  \bibAnnoteFile{NoStop}{Grubmueller1995}%
\bibitem{Voter1997b}%
  \BibitemOpen
  \bibfield{author}{%
  \bibinfo {author} {\bibfnamefont{A.~F.}\ \bibnamefont{Voter}},\ }%
  \bibfield{journal}{%
  \bibinfo {journal} {Phys. Rev. Lett.}\ }%
  \textbf{\bibinfo {volume} {78}},\ \bibinfo {pages} {3908} (\bibinfo {year}
  {1997})%
  \bibAnnoteFile{NoStop}{Voter1997b}%
\bibitem{WuW1999}%
  \BibitemOpen
  \bibfield{author}{%
  \bibinfo {author} {\bibfnamefont{X.}~\bibnamefont{Wu}}\ and\ \bibinfo
  {author} {\bibfnamefont{S.}~\bibnamefont{Wang}},\ }%
  \bibfield{journal}{%
  \bibinfo {journal} {J. Chem. Phys.}\ }%
  \textbf{\bibinfo {volume} {110}},\ \bibinfo {pages} {9401} (\bibinfo {year}
  {1999})%
  \bibAnnoteFile{NoStop}{WuW1999}%
\bibitem{YanP1999}%
  \BibitemOpen
  \bibfield{author}{%
  \bibinfo {author} {\bibfnamefont{Q.}~\bibnamefont{Yan}}\ and\ \bibinfo
  {author} {\bibfnamefont{J.~J.}\ \bibnamefont{de~Pablo}},\ }%
  \bibfield{journal}{%
  \bibinfo {journal} {J. Chem. Phys.}\ }%
  \textbf{\bibinfo {volume} {111}},\ \bibinfo {pages} {9509} (\bibinfo {year}
  {1999})%
  \bibAnnoteFile{NoStop}{YanP1999}%
\bibitem{MitsutakeSO2001}%
  \BibitemOpen
  \bibfield{author}{%
  \bibinfo {author} {\bibfnamefont{A.}~\bibnamefont{Mitsutake}}, \bibinfo
  {author} {\bibfnamefont{Y.}~\bibnamefont{Sugita}},\ and\ \bibinfo {author}
  {\bibfnamefont{Y.}~\bibnamefont{Okamoto}},\ }%
  \bibfield{journal}{%
  \bibinfo {journal} {Biopolymers (Peptide Sci.)}\ }%
  \textbf{\bibinfo {volume} {60}},\ \bibinfo {pages} {96} (\bibinfo {year}
  {2001})%
  \bibAnnoteFile{NoStop}{MitsutakeSO2001}%
\bibitem{laio2002escaping}%
  \BibitemOpen
  \bibfield{author}{%
  \bibinfo {author} {\bibfnamefont{A.}~\bibnamefont{Laio}}\ and\ \bibinfo
  {author} {\bibfnamefont{M.}~\bibnamefont{Parrinello}},\ }%
  \bibfield{journal}{%
  \bibinfo {journal} {Proc. Natl. Acad. Sci. U.S.A.}\ }%
  \textbf{\bibinfo {volume} {99}},\ \bibinfo {pages} {12562} (\bibinfo {year}
  {2002})%
  \bibAnnoteFile{NoStop}{laio2002escaping}%
\bibitem{AndricioaeiDK2003}%
  \BibitemOpen
  \bibfield{author}{%
  \bibinfo {author} {\bibfnamefont{I.}~\bibnamefont{Andricioaei}}, \bibinfo
  {author} {\bibfnamefont{A.~R.}\ \bibnamefont{Dinner}},\ and\ \bibinfo
  {author} {\bibfnamefont{M.}~\bibnamefont{Karplus}},\ }%
  \bibfield{journal}{%
  \bibinfo {journal} {J. Chem. Phys.}\ }%
  \textbf{\bibinfo {volume} {118}},\ \bibinfo {pages} {1074} (\bibinfo {year}
  {2003})%
  \bibAnnoteFile{NoStop}{AndricioaeiDK2003}%
\bibitem{HamelbergSM2005b}%
  \BibitemOpen
  \bibfield{author}{%
  \bibinfo {author} {\bibfnamefont{D.}~\bibnamefont{Hamelberg}}, \bibinfo
  {author} {\bibfnamefont{T.-Y.}\ \bibnamefont{Shen}},\ and\ \bibinfo {author}
  {\bibfnamefont{J.~A.}\ \bibnamefont{McCammon}},\ }%
  \bibfield{journal}{%
  \bibinfo {journal} {J. Chem. Phys.}\ }%
  \textbf{\bibinfo {volume} {122}},\ \bibinfo {pages} {241103} (\bibinfo {year}
  {2005})%
  \bibAnnoteFile{NoStop}{HamelbergSM2005b}%
\bibitem{ZhouJKZR2006}%
  \BibitemOpen
  \bibfield{author}{%
  \bibinfo {author} {\bibfnamefont{X.}~\bibnamefont{Zhou}}, \bibinfo {author}
  {\bibfnamefont{Y.}~\bibnamefont{Jiang}}, \bibinfo {author}
  {\bibfnamefont{K.}~\bibnamefont{Kremer}}, \bibinfo {author}
  {\bibfnamefont{H.}~\bibnamefont{Ziock}},\ and\ \bibinfo {author}
  {\bibfnamefont{S.}~\bibnamefont{Rasmussen}},\ }%
  \bibfield{journal}{%
  \bibinfo {journal} {Phys. Rev. E}\ }%
  \textbf{\bibinfo {volume} {74}},\ \bibinfo {pages} {R035701} (\bibinfo {year}
  {2006})%
  \bibAnnoteFile{NoStop}{ZhouJKZR2006}%
\bibitem{barducci2008well}%
  \BibitemOpen
  \bibfield{author}{%
  \bibinfo {author} {\bibfnamefont{A.}~\bibnamefont{Barducci}}, \bibinfo
  {author} {\bibfnamefont{G.}~\bibnamefont{Bussi}},\ and\ \bibinfo {author}
  {\bibfnamefont{M.}~\bibnamefont{Parrinello}},\ }%
  \bibfield{journal}{%
  \bibinfo {journal} {Phys. Rev. Lett.}\ }%
  \textbf{\bibinfo {volume} {100}},\ \bibinfo {pages} {020603} (\bibinfo {year}
  {2008})%
  \bibAnnoteFile{NoStop}{barducci2008well}%
\bibitem{wang2001efficient}%
  \BibitemOpen
  \bibfield{author}{%
  \bibinfo {author} {\bibfnamefont{F.}~\bibnamefont{Wang}}\ and\ \bibinfo
  {author} {\bibfnamefont{D.~P.}\ \bibnamefont{Landau}},\ }%
  \bibfield{journal}{%
  \bibinfo {journal} {Phys. Rev. Lett.}\ }%
  \textbf{\bibinfo {volume} {86}},\ \bibinfo {pages} {2050} (\bibinfo {year}
  {2001})%
  \bibAnnoteFile{NoStop}{wang2001efficient}%
\bibitem{gong2009structuring}%
  \BibitemOpen
  \bibfield{author}{%
  \bibinfo {author} {\bibfnamefont{L.}~\bibnamefont{Gong}}\ and\ \bibinfo
  {author} {\bibfnamefont{X.}~\bibnamefont{Zhou}},\ }%
  \bibfield{journal}{%
  \bibinfo {journal} {Phys. Rev. E}\ }%
  \textbf{\bibinfo {volume} {80}},\ \bibinfo {pages} {026707} (\bibinfo {year}
  {2009})%
  \bibAnnoteFile{NoStop}{gong2009structuring}%
\bibitem{ZhangM2009}%
  \BibitemOpen
  \bibfield{author}{%
  \bibinfo {author} {\bibfnamefont{C.}~\bibnamefont{Zhang}}\ and\ \bibinfo
  {author} {\bibfnamefont{J.}~\bibnamefont{Ma}},\ }%
  \bibfield{journal}{%
  \bibinfo {journal} {J. Chem. Phys.}\ }%
  \textbf{\bibinfo {volume} {130}},\ \bibinfo {pages} {194112} (\bibinfo {year}
  {2009})%
  \bibAnnoteFile{NoStop}{ZhangM2009}%
\bibitem{XuZJ2015}%
  \BibitemOpen
  \bibfield{author}{%
  \bibinfo {author} {\bibfnamefont{S.}~\bibnamefont{Xu}}, \bibinfo {author}
  {\bibfnamefont{X.}~\bibnamefont{Zhou}}, \bibinfo {author}
  {\bibfnamefont{Y.}~\bibnamefont{Jiang}},\ and\ \bibinfo {author}
  {\bibfnamefont{Y.}~\bibnamefont{Wang}},\ }%
  \bibfield{journal}{%
  \bibinfo {journal} {Sci. China-Phys. Mech. Astron.}\ }%
  \textbf{\bibinfo {volume} {58}},\ \bibinfo {pages} {090501} (\bibinfo {year}
  {2015})%
  \bibAnnoteFile{NoStop}{XuZJ2015}%
\bibitem{jarzynski1997nonequilibrium}%
  \BibitemOpen
  \bibfield{author}{%
  \bibinfo {author} {\bibfnamefont{C.}~\bibnamefont{Jarzynski}},\ }%
  \bibfield{journal}{%
  \bibinfo {journal} {Phys. Rev. Lett.}\ }%
  \textbf{\bibinfo {volume} {78}},\ \bibinfo {pages} {2690} (\bibinfo {year}
  {1997})%
  \bibAnnoteFile{NoStop}{jarzynski1997nonequilibrium}%
\bibitem{PhysRevE.92.012131}%
  \BibitemOpen
  \bibfield{author}{%
  \bibinfo {author} {\bibfnamefont{Z.}~\bibnamefont{Gong}}\ and\ \bibinfo
  {author} {\bibfnamefont{H.~T.}\ \bibnamefont{Quan}},\ }%
  \bibfield{journal}{%
  \Doi{10.1103/PhysRevE.92.012131}{\bibinfo {journal} {Phys. Rev. E}}\ }%
  \textbf{\bibinfo {volume} {92}},\ \bibinfo {pages} {012131} (\bibinfo {year}
  {2015})%
  \bibAnnoteFile{NoStop}{PhysRevE.92.012131}%
\bibitem{park2003free}%
  \BibitemOpen
  \bibfield{author}{%
  \bibinfo {author} {\bibfnamefont{S.}~\bibnamefont{Park}}, \bibinfo {author}
  {\bibfnamefont{F.}~\bibnamefont{Khalili-Araghi}}, \bibinfo {author}
  {\bibfnamefont{E.}~\bibnamefont{Tajkhorshid}},\ and\ \bibinfo {author}
  {\bibfnamefont{K.}~\bibnamefont{Schulten}},\ }%
  \bibfield{journal}{%
  \bibinfo {journal} {J. Chem. Phys.}\ }%
  \textbf{\bibinfo {volume} {119}},\ \bibinfo {pages} {3559} (\bibinfo {year}
  {2003})%
  \bibAnnoteFile{NoStop}{park2003free}%
\bibitem{liphardt2002equilibrium}%
  \BibitemOpen
  \bibfield{author}{%
  \bibinfo {author} {\bibfnamefont{J.}~\bibnamefont{Liphardt}}, \bibinfo
  {author} {\bibfnamefont{S.}~\bibnamefont{Dumont}}, \bibinfo {author}
  {\bibfnamefont{S.~B.}\ \bibnamefont{Smith}}, \bibinfo {author}
  {\bibfnamefont{I.}~\bibnamefont{Tinoco}},\ and\ \bibinfo {author}
  {\bibfnamefont{C.}~\bibnamefont{Bustamante}},\ }%
  \bibfield{journal}{%
  \bibinfo {journal} {Science}\ }%
  \textbf{\bibinfo {volume} {296}},\ \bibinfo {pages} {1832} (\bibinfo {year}
  {2002})%
  \bibAnnoteFile{NoStop}{liphardt2002equilibrium}%
\bibitem{hummer2001free}%
  \BibitemOpen
  \bibfield{author}{%
  \bibinfo {author} {\bibfnamefont{G.}~\bibnamefont{Hummer}}\ and\ \bibinfo
  {author} {\bibfnamefont{A.}~\bibnamefont{Szabo}},\ }%
  \bibfield{journal}{%
  \bibinfo {journal} {Proc. Natl. Acad. Sci. U.S.A.}\ }%
  \textbf{\bibinfo {volume} {98}},\ \bibinfo {pages} {3658} (\bibinfo {year}
  {2001})%
  \bibAnnoteFile{NoStop}{hummer2001free}%
\bibitem{ChuanBiao}%
  \BibitemOpen
  \bibfield{author}{%
  \bibinfo {author} {\bibfnamefont{Z.}~\bibnamefont{Chuan-Biao}}, \bibinfo
  {author} {\bibfnamefont{L.}~\bibnamefont{Ming}},\ and\ \bibinfo {author}
  {\bibfnamefont{Z.}~\bibnamefont{Xin}},\ }%
  \bibfield{journal}{%
  \Doi{10.1088/1674-1056/24/12/120202}{\bibinfo {journal} {Chin. Phys. B}}\ }%
  \textbf{\bibinfo {volume} {24}},\ \bibinfo {eid} {120202} (\bibinfo {year}
  {2015})%
  \bibAnnoteFile{NoStop}{ChuanBiao}%
\bibitem{GongZO2015}%
  \BibitemOpen
  \bibfield{author}{%
  \bibinfo {author} {\bibfnamefont{L.}~\bibnamefont{Gong}}, \bibinfo {author}
  {\bibfnamefont{X.}~\bibnamefont{Zhou}},\ and\ \bibinfo {author}
  {\bibfnamefont{Z.-C.}\ \bibnamefont{OuYang}},\ }%
  \bibfield{journal}{%
  \bibinfo {journal} {Chin. Phys. B}\ }%
  \textbf{\bibinfo {volume} {24}},\ \bibinfo {pages} {060202} (\bibinfo {year}
  {2015})%
  \bibAnnoteFile{NoStop}{GongZO2015}%
\bibitem{hummer2005free}%
  \BibitemOpen
  \bibfield{author}{%
  \bibinfo {author} {\bibfnamefont{G.}~\bibnamefont{Hummer}}\ and\ \bibinfo
  {author} {\bibfnamefont{A.}~\bibnamefont{Szabo}},\ }%
  \bibfield{journal}{%
  \bibinfo {journal} {Acc. Chem. Res.}\ }%
  \textbf{\bibinfo {volume} {38}},\ \bibinfo {pages} {504} (\bibinfo {year}
  {2005})%
  \bibAnnoteFile{NoStop}{hummer2005free}%
\bibitem{zuckerman2002theory}%
  \BibitemOpen
  \bibfield{author}{%
  \bibinfo {author} {\bibfnamefont{D.~M.}\ \bibnamefont{Zuckerman}}\ and\
  \bibinfo {author} {\bibfnamefont{T.~B.}\ \bibnamefont{Woolf}},\ }%
  \bibfield{journal}{%
  \bibinfo {journal} {Phys. Rev. Lett.}\ }%
  \textbf{\bibinfo {volume} {89}},\ \bibinfo {pages} {180602} (\bibinfo {year}
  {2002})%
  \bibAnnoteFile{NoStop}{zuckerman2002theory}%
\bibitem{gore2003bias}%
  \BibitemOpen
  \bibfield{author}{%
  \bibinfo {author} {\bibfnamefont{J.}~\bibnamefont{Gore}}, \bibinfo {author}
  {\bibfnamefont{F.}~\bibnamefont{Ritort}},\ and\ \bibinfo {author}
  {\bibfnamefont{C.}~\bibnamefont{Bustamante}},\ }%
  \bibfield{journal}{%
  \bibinfo {journal} {Proc. Natl. Acad. Sci. U.S.A.}\ }%
  \textbf{\bibinfo {volume} {100}},\ \bibinfo {pages} {12564} (\bibinfo {year}
  {2003})%
  \bibAnnoteFile{NoStop}{gore2003bias}%
\bibitem{park2004calculating}%
  \BibitemOpen
  \bibfield{author}{%
  \bibinfo {author} {\bibfnamefont{S.}~\bibnamefont{Park}}\ and\ \bibinfo
  {author} {\bibfnamefont{K.}~\bibnamefont{Schulten}},\ }%
  \bibfield{journal}{%
  \bibinfo {journal} {J. Chem. Phys.}\ }%
  \textbf{\bibinfo {volume} {120}},\ \bibinfo {pages} {5946} (\bibinfo {year}
  {2004})%
  \bibAnnoteFile{NoStop}{park2004calculating}%
\bibitem{lammps}%
  \BibitemOpen
  \bibinfo {journal} {See http://lammps.sandia.gov/ for information about the
  LAMMPS molecular dynamics simulator.}%
  \bibAnnoteFile{Stop}{lammps}%
\bibitem{hansen1969phase}%
  \BibitemOpen
\bibfield{journal}{%
    }%
  \bibfield{author}{%
  \bibinfo {author} {\bibfnamefont{J.-P.}\ \bibnamefont{Hansen}}\ and\ \bibinfo
  {author} {\bibfnamefont{L.}~\bibnamefont{Verlet}},\ }%
  \bibfield{journal}{%
  \bibinfo {journal} {Phys. Rev.}\ }%
  \textbf{\bibinfo {volume} {184}},\ \bibinfo {pages} {151} (\bibinfo {year}
  {1969})%
  \bibAnnoteFile{NoStop}{hansen1969phase}%
\bibitem{steinhardt1983bond}%
  \BibitemOpen
  \bibfield{author}{%
  \bibinfo {author} {\bibfnamefont{P.~J.}\ \bibnamefont{Steinhardt}}, \bibinfo
  {author} {\bibfnamefont{D.~R.}\ \bibnamefont{Nelson}},\ and\ \bibinfo
  {author} {\bibfnamefont{M.}~\bibnamefont{Ronchetti}},\ }%
  \bibfield{journal}{%
  \bibinfo {journal} {Phys. Rev. B}\ }%
  \textbf{\bibinfo {volume} {28}},\ \bibinfo {pages} {784} (\bibinfo {year}
  {1983})%
  \bibAnnoteFile{NoStop}{steinhardt1983bond}%
\bibitem{lechner2008accurate}%
  \BibitemOpen
  \bibfield{author}{%
  \bibinfo {author} {\bibfnamefont{W.}~\bibnamefont{Lechner}}\ and\ \bibinfo
  {author} {\bibfnamefont{C.}~\bibnamefont{Dellago}},\ }%
  \bibfield{journal}{%
  \bibinfo {journal} {J. Chem. Phys.}\ }%
  \textbf{\bibinfo {volume} {129}},\ \bibinfo {pages} {114707} (\bibinfo {year}
  {2008})%
  \bibAnnoteFile{NoStop}{lechner2008accurate}%
\bibitem{russo2012microscopic}%
  \BibitemOpen
  \bibfield{author}{%
  \bibinfo {author} {\bibfnamefont{J.}~\bibnamefont{Russo}}\ and\ \bibinfo
  {author} {\bibfnamefont{H.}~\bibnamefont{Tanaka}},\ }%
  \bibfield{journal}{%
  \bibinfo {journal} {Sci. Rep.}\ }%
  \textbf{\bibinfo {volume} {2}} (\bibinfo {year} {2012})%
  \bibAnnoteFile{NoStop}{russo2012microscopic}%
\end{thebibliography}%
 

\end{document}